\documentclass[aip,apl,reprint,amsmath,amssymb,amsfonts,floatfix,intlimits,twocolumn]{revtex4-2}

\usepackage{graphicx}
\usepackage{setspace}
\usepackage{bm}

\def\<{\left\langle}
\def\>{\right\rangle}

\begin{document}

\title
{
RF signal detector and energy harvester based on a spin-torque diode with perpendicular magnetic anisotropy
}

\author{P.Yu.~Artemchuk}

\affiliation{Faculty of Radio Physics, Electronics and Computer Systems,
Taras Shevchenko National University of Kyiv, Kyiv 01601, Ukraine}
\affiliation{Department of Physics, Oakland University, Rochester, Michigan 48309, USA}

\author{O.V.~Prokopenko}
\email{Oleksandr.Prokopenko@gmail.com}
\affiliation{Faculty of Radio Physics, Electronics and Computer Systems,
Taras Shevchenko National University of Kyiv, Kyiv 01601, Ukraine}

\author{E.N.~Bankowski}
\author{T.J.~Meitzler}
\affiliation{U.S. Army CCDC GVSC, Warren, Michigan 48397, USA}

\author{V.S.~Tyberkevych}
\author{A.N.~Slavin}
\affiliation{Department of Physics, Oakland University, Rochester, Michigan 48309, USA}

\begin{abstract}

We demonstrate theoretically that in a spintronic diode (SD), having a free magnetic layer with perpendicular magnetic anisotropy of the first and second order and no external bias magnetic field, the out-of-plane regime of magnetization precession can be excited by sufficiently large (exceeding a certain threshold) RF signals  with the frequencies $\lesssim 250$~MHz. We also show that such a device can operate as a broadband energy harvester capable of converting incident RF power into a DC power with the conversion efficiency of $\sim5$\%. The developed analytical theory of the bias-free SD operation can be used for the optimization of high-efficiency RF detectors and energy harvesters based on SDs.
\end{abstract}

\maketitle


With the advances of the Internet of Things (IOT) and radio frequency identification (RFID) technologies and wide application of micro- and nano-scale wireless devices that require independent power supplies, the problems of efficient RF signal detection and energy harvesting from ambient sources of radiation became critically important
\cite{Harb2011RenEn, Duroc2014Book, Valenta2014MMag, Divakaran2019MCAE, Pozo2019Electronics}.
Both these problems can be solved with the help of spintronic diodes (SDs) based on magnetic tunnel junctions (MTJs)
\cite{Tulapurkar2005Nature,
Prokopenko2013Book,
Prokopenko2015LTP,
Ishibashi2010APEx,
Prokopenko2011APL,
Miwa2014NatMater,
Fang2016NatCommun,
Cheng2010PRL,
Prokopenko2012JAP,
Fang2019PRAppl,
Tomasello2020PRAppl,
Artemchuk2020Book}.
In such a diode, the input RF current $I(t) = I_{\rm RF} \cos(\omega t)$ applied to an MTJ excites variations of the junction's resistance $R(t)$ with the angular frequency $\omega = 2 \pi f$ of the incident RF signal.
As a result, the voltage across the MTJ has a DC component
$U_{\rm DC} = \<I(t)R(t)\>$ (here the angular brackets denote averaging over the period $T = 2\pi / \omega = 1/f$ of the RF current, $\<x\> = T^{-1} \int_0^T x \, dt$) \cite{Tulapurkar2005Nature, Prokopenko2013Book, Prokopenko2015LTP, Artemchuk2020Book}.

There are several possible regimes of operation of a SD \cite{Prokopenko2013Book, Prokopenko2015LTP, Artemchuk2020Book}.
Among them, the best-known regime is the regime of the forced in-plane (IP) magnetization precession about the in-plane ``easy axis'' in the free layer (FL) of the SD.
This regime is, sometimes, called a ``resonance'' regime, because the DC voltage produced by a SD has a maximum magnitude when the driving external RF signal has the frequency equal to the frequency of the ferromagnetic resonance of the SD's FL \cite{Tulapurkar2005Nature, Prokopenko2013Book, Prokopenko2015LTP, Ishibashi2010APEx, Prokopenko2011APL,Miwa2014NatMater,Fang2016NatCommun, Artemchuk2020Book}.

Another regime, which exists in the case when an SD is biased by a perpendicular bias magnetic field insufficient for the full saturation of the SD FL, is the regime of the forced out-of-plane (OOP) precession, and it was, first, described theoretically in \cite{Prokopenko2012JAP}.
In the OOP-regime, the equilibrium direction of the FL magnetization lies out-of-plane of the SD FL, and the incident RF signal (if it exceeds a certain amplitude threshold) excites in the SD FL a large-angle OOP magnetization precession about the perpendicular direction of the bias magnetic field \cite{Prokopenko2012JAP, Prokopenko2013Book, Prokopenko2015LTP}.
In this regime the magnitude of the SD output DC voltage $U_{\rm DC}$ is negligibly small in the case when the incident RF current amplitude $I_{\rm RF}$ is below a certain threshold $I_{\rm th\,}$, but increases abruptly as soon as $I_{\rm RF\,}$ exceeds certain threshold value $I_{\rm th\,}$.
Above the threshold, the angle of the OOP magnetization precession is only slowly increasing with the increase of the  $I_{\rm RF}$, and the resultant DC voltage produced by the SD virtually does not depend on the magnitude of the driving RF signal.  This DC voltage also depends on the driving frequency in a non-resonance way -- it increases with the increase of the driving frequency up to a certain magnitude, and for higher frequencies vanishes abruptly \cite{Prokopenko2012JAP, Prokopenko2013Book, Prokopenko2015LTP}. Thus, the SD in the OOP regime works as a non-resonant threshold detector of RF signals having a sufficiently low frequency \cite{Prokopenko2012JAP, Prokopenko2013Book, Prokopenko2015LTP}.

We believe that the OOP regime of the SD operation was observed for the first time in \cite{Cheng2010PRL}, and is responsible for the extremely large diode volt-watt sensitivity observed in the experiments \cite{Cheng2010PRL}.
It was also proposed in  \cite{Prokopenko2012JAP, Prokopenko2013Book, Prokopenko2015LTP} that the OOP regime of the SD operation could be very useful for the broadband RF energy harvesting.

Recent experiments \cite{Fang2019PRAppl}, indeed, demonstrated that the efficient broadband RF energy harvesting is possible using an SD working in the OOP regime, and an important practical achievement of \cite{Fang2019PRAppl} was the demonstration that DC energy harvesting in the OOP regime can be experimentally realized \emph{without any bias magnetic field}.
To move the equilibrium orientation of the FL magnetization out of plane of the SD FL the authors of \cite{Fang2019PRAppl} used an FL with \emph{perpendicular magnetic anisotropy} (PMA).

The use of PMA in the forced RF dynamics of a SD creates a new situation, which was not previously considered theoretically, and in our current work we consider both analytically and numerically the OOP regime of the SD operation in the case when the out-of-plane equilibrium orientation of the static magnetization of the SD FL was created by the combination of the FL demagnetization field and the FL PMA.
The goal is to elucidate the forced RF magnetization dynamics in this case to be able to optimize the operation of the broadband RF energy harvesters based on SDs operation in the OOP regime.


\begin{figure}
\includegraphics{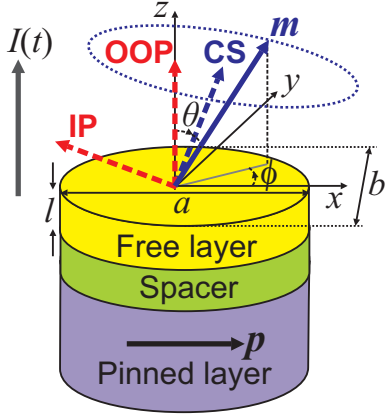}
\caption
{
(Color online)
Model of a considered spintronic diode (SD) with a free layer (FL) having perpendicular magnetic anisotropy (PMA). Due to the competition between the demagnetization field and the PMA of the FL its static magnetization has the out-of-plane (OOP) orientation (dashed blue line corresponding to the equilibrium cone state (CS) of the magnetization). Under the action of an external RF current $I(t)$ the unit magnetization vector ${\bm m}$ of the FL (solid blue arrow) starts to precess along the OOP trajectory (dashed blue curve) about the CS direction. The directions of the static magnetization for the other two possible equilibrium states (the perpendicular OOP and the in-plane (IP) states) are shown by red dashed arrows.
}
\label{f:Model}
\end{figure}

We consider an SD formed by an MTJ nano-pillar having elliptical-shape $a \times b$ FL of the thickness $l$ ($a/2$, and $b/2$ are the ellipse semi-axes, Fig.~\ref{f:Model}).
We assume that the magnetization of the FL ${\bm M} = {\bm m} M_{s\,}$ is spatially uniform, and depends on time $t$ only (i.e., we use the macrospin approximation \cite{Macrospin}), ${\bm m} \equiv {\bm m}(t)$ is the unit vector, and $M_s$ is the saturation magnetization of the FL.
The magnetization of the lowest pinned layer is assumed to be completely fixed, and its direction is defined by the unit vector ${\bm p} = {\hat{\bm x}}$, where ${\hat{\bm x}}$ is the unit vector of the \emph{x}-axis.
The FL of the SD (see Fig.~\ref{f:Model}) has a PMA of the first and second order \cite{Fang2019PRAppl}, characterized by the anisotropy constants $K_1$ and $K_2$, respectively, and no in-plane anisotropy.
There is no bias magnetic field applied to the structure.

Then, the effective magnetic field ${\bm B}_{\rm eff}$ acting on the FL magnetization ${\bm M}$ is formed by the demagnetization field \cite{Melkov1996Book} ${\bm B}_{\rm d} = -\mu_0 M_s \left({\bm N} \cdot {\bm m}\right)$ and the PMA field \cite{Fang2019PRAppl} ${\bm B}_{\rm PMA} = {\hat{\bm z}}\left[B_1 + B_2 \left(1-m_z^2\right)\right] m_z$,
${\bm B}_{\rm eff} = {\bm B}_{\rm d} + {\bm B}_{\rm PMA}$.
Here ${\bm N} = {\rm diag}\{N_x, N_y, N_z\}$ is the diagonal self-demagnetization tensor having the elements $N_x$, $N_y$, and $N_z$ (their sum is equal to 1),
${\hat{\bm z}}$ is the unit vector of \emph{z}-axis, $m_z = {\bm m} \cdot {\hat{\bm z}}$, $\mu_0$ is the vacuum permeability ,$B_1 = 2 K_1 / M_s$ and $B_2 = 4 K_2 / M_s$ are the fields of the first- and second-order PMA, respectively.

The magnetization dynamics in the FL is governed by the Landau-Lifshitz-Gilbert-Slonczewski equation \cite{Prokopenko2013Book, Prokopenko2015LTP, Slavin2009IEEETMagn}:
\begin{eqnarray}
\label{eq:LLGS}
	\frac{d{\bm m}}{dt}
    &=&
    \gamma\left[{\bm B}_{\rm eff} \times {\bm m}\right]
	+
    \alpha \left[{\bm m} \times \frac{d{\bm m}}{dt}\right]
    +
    \nonumber\\
    &&
	\sigma I(t)
    \left[{\bm m} \times \left[{\bm m} \times {\bm p}\,\right]\right]
    \ ,
\end{eqnarray}
where $\gamma \approx 2\pi \cdot 28$~GHz/T is the modulus of the gyromagnetic ratio,
$\alpha$ is the Gilbert damping constant,
$\sigma = \sigma_{\bot}/(1 + \eta^2\cos\beta)$ is the current-torque proportionality coefficient,
$\sigma_{\bot} = (\gamma\hbar/2e)\eta/(M_s V)$,
$\hbar$ is the reduced Planck constant,
$e$ is the modulus of the electron charge,
$\eta$ is the spin-polarization of current,
$V = \pi a b l / 4$ is the volume of the FL having thickness $l$ and elliptical cross section $a \times b$,
and $\beta = \arccos({\bm m} \cdot {\bm p})$ is the angle between the magnetizations of the free and pinned layers.

There are three possible equilibrium magnetization states in the considered SD (see Fig.~\ref{f:Model}): the OOP state (${\bm m} = \pm {\hat{\bm z}}$,
$m_z = \pm 1$), the IP state (${\bm m}$ lies in the \emph{x}-\emph{y} plane, $m_z = 0$), and the cone state (CS) that corresponds to the case $0 < |m_z| < 1$.
Among these three possible equilibrium magnetization states, the CS is the most interesting for broadband RF signal detection and RF energy harvesting, as in this equilibrium state an external RF signal most easily excites in the SD FL the magnetization precession with a large precession angle. In the absence of the bias magnetic field, the SD dynamics is symmetrical with respect to $180^\circ$ rotation around $x$ axis; therefore, we shall consider below only the case when the equilibrium magnetization direction lies in the upper half sphere, $m_z > 0$.

The value of $m_z$ that corresponds to the equilibrium CS of magnetization,
$m_{z, {\rm CS}}$, can be found from (\ref{eq:LLGS}) assuming $d{\bm m}/dt = 0$, $I(t) = 0$:
$m_{z, {\rm CS}} = \sqrt{1 - r}$, where $r = [\mu_0 M_s (3 N_z - 1)/2 - B_1] / B_2$ is the dimensionless ratio describing the state of the considered system.
The equilibrium magnetization angle $\theta_{\rm CS}$ ($m_z = \cos\theta$) that corresponds to the CS is
$\theta_{\rm CS} = \arccos(\sqrt{1 - r})$.

It is clear, that the equilibrium CS of magnetization is possible only when $0 < r < 1$, i.e., when the second-order PMA field $B_2$ is stronger than the positively-definite effective first-order field $[\mu_0 M_s (3 N_z - 1)/2 - B_1] > 0$. Alternatively, this condition can be written as a constraint on allowed values of $N_z$: $1 + 2 B_1 / \mu_0 M_s < 3 N_z < 1 + 2 (B_1 + B_2)/ \mu_0 M_s$.
When $r \rightarrow 0$, the CS transforms to the OOP state ($m_z \rightarrow 1$), while at $r \rightarrow 1$ the CS turns into the IP state ($m_z \rightarrow 0$).

Using spherical polar coordinates for the unit vector ${\bm m}$,
${\bm m} = {\hat{\bm x}} \sin\theta \cos\phi +
{\hat{\bm y}} \sin\theta \sin\phi + {\hat{\bm z}} \cos\theta$
(where ${\hat{\bm y}}$ is the unit vector of \emph{y}-axis), one can derive from (\ref{eq:LLGS}) equations for the polar $\theta \equiv \theta(t)$ and azimuthal $\phi \equiv \phi(t)$ magnetization angles:
\begin{subequations}
\label{eq:ThetaPhi}
\begin{eqnarray}
\label{eq:ThetaPhi-Theta}
    \frac{d \theta}{d t}
    &=&
    -
    \Bigl[
        \alpha \omega_\theta \sin\theta
        +
        \sigma I(t)
        \left(
            \cos\theta \cos\phi
            +
            \alpha \sin\phi
        \right)
        +
    \nonumber\\
    &&
        \alpha \omega_M \sin\theta \cos\theta
        \left(N_x \cos^2\phi + N_y \sin^2\phi\right)
        -
    \nonumber\\
    &&
        \omega_M (N_x - N_y) \sin\theta \sin\phi \cos\phi
    \Bigr]
    /
    \left(1 + \alpha^2\right)
    \ , \ \ \\
\label{eq:ThetaPhi-Phi}
	\frac{d \phi}{d t}
    &=&
    \Bigl[
        \omega_\theta
        -
        \sigma I(t)
        \left(
            \alpha \cot\theta \cos\phi
            -
            \csc\theta \sin\phi
        \right)
        +
    \nonumber\\
    &&
        \omega_M \cos\theta \left(N_x \cos^2\phi + N_y \sin^2\phi\right)
        +
    \nonumber\\
    &&
        \alpha \omega_M (N_x - N_y) \sin\phi \cos\phi
    \Bigr]
    /
    \left(1 + \alpha^2\right)
    \ .
\end{eqnarray}
\end{subequations}
Here
$\omega_\theta \equiv \omega_\theta(\theta) =
(\omega_2 \sin^{2}\theta + \omega_1 - \omega_M N_z) \cos\theta$,
$\omega_1 = \gamma B_{1\,}$,
$\omega_2 = \gamma B_{2\,}$,
$\omega_M = \gamma \mu_0 M_{s\,}$.
Assuming that both the Gilbert damping constant $\alpha$ and the magnitude of the input RF current $I_{\rm RF}$ are rather small, we can substantially simplify (\ref{eq:ThetaPhi}) by neglecting terms proportional to $\alpha^2$ and $\alpha I_{\rm RF}$. Note, however, that this approximation should not be used for the case of large-power input signals ($I_{\rm RF} \gg I_{\rm th}$) \cite{Tomasello2020PRAppl}.

To estimate the average influence of an input RF current on the magnetization dynamics, we assume that in the OOP-regime the magnetization precesses around some equilibrium inclined axis (corresponding to the equilibrium CS with the polar angle $\theta_{\rm CS}$) along an approximately circular orbit (see Fig.~\ref{f:Model}).
First, we let $\theta \approx {\rm const}$, $\phi \approx \omega t + \psi$ in the CS, where $\psi$ is the phase shift between the magnetization oscillations and the driving current.
Second, we average the simplified equations for $\theta$ and $\phi$ over the period of oscillation $T = 2\pi/\omega$ of the driving current, and obtain the following equations for the slow variables $\theta$ and $\psi$:
\begin{subequations}
\label{eq:ThetaPsiAvg}
\begin{eqnarray}
\label{eq:ThetaPsiAvg-Theta}
    \< \frac{d \theta}{d t} \>
    &=&
    - \alpha \omega_p \sin\theta
    -
    v \frac{\sigma_\bot I_{\rm RF}}{2} \cos\theta \cos\psi
    \, , \\
\label{eq:ThetaPsiAvg-Psi}
	\< \frac{d \psi}{d t} \>
    &=&
    \omega_p - \omega
    +
    u \frac{\sigma_\bot I_{\rm RF}}{2} \frac{1}{\sin\theta} \sin\psi
    \ .
\end{eqnarray}
\end{subequations}
Here we used the relation $N_x + N_y = 1 - N_z$ and introduced the frequency of the OOP precession $\omega_p \equiv \omega_p(\theta) = \omega_\theta + \omega_M \cos\theta (1-N_z)/2 =\omega_2 \left(\sin^2\theta - r\right) \cos\theta$ and the dimensionless functions $u \equiv u(a_\eta) = \left[1 - (q_\eta - 1)^2 / a^2_\eta \right]/q_\eta$, $v \equiv v(a_\eta) = \left[1 + (q_\eta - 1)^2 / a^2_\eta \right]/q_\eta$ of parameter $a_\eta = \eta^2 \sin\theta$, $q_\eta = \sqrt{1-a^2_\eta}$. In a typical experimental situation ($\eta \le 0.7$) the values of $u$ and $v$ are close to 1 for all the angles $0 \le \theta \le \pi/2$.

The OOP-regime of magnetization precession corresponds to the following stationary solution of (\ref{eq:ThetaPsiAvg}): $\theta = \theta_s = {\rm const}$, $\psi = \psi_s = {\rm const}$.
In this case one can find from (\ref{eq:ThetaPsiAvg}) the stationary value of the phase shift $\psi_{s\,}$: $\sin\psi_s = (2 / u) (\omega - \omega_p) \sin\theta_s / \sigma_{\bot} I_{\rm RF\,}$,
$\cos\psi_s = -2(\alpha / v) \omega_p \tan\theta_s / \sigma_{\bot} I_{\rm RF\,}$,
and, then, obtain the characteristic equation for the stationary polar precession angle $\theta_s$:
\begin{equation}
\label{eq:CharEq}
    \left( \omega - \omega_p \right)^2 \sin^2\theta_s
    +
    \frac{u^2}{v^2} \alpha^2 \omega_p^2 \tan^2\theta_s
    =
    \frac{u^2}{4} \sigma^2_{\bot} I^2_{\rm RF}
    \ .
\end{equation}

Eq.~(\ref{eq:CharEq}) has solutions only for RF currents $I_{\rm RF}$ that are larger than a certain threshold $I_{\rm th}$.
For small damping ($\alpha \ll 1$) the first term in (\ref{eq:CharEq}) is much larger than the second one unless $\omega_p \approx \omega$. Then, we can assume that at the threshold $I_{\rm RF} = I_{\rm th}$ the OOP eigen-frequency $\omega_p(\theta)$ coincides with the driving frequency $\omega$, which determines the threshold precession angle $\theta_{\rm th}$: $\omega_p(\theta_{\rm th}) = \omega$. Using this angle in (\ref{eq:CharEq}), one can obtain the following expression for the threshold current:
\begin{equation}
\label{eq:Ith}
    I_{\rm th}
    \approx
    2
    \frac{\alpha}{v}
    \frac{\omega}{\sigma_\bot}
    \tan\theta_{\rm th}
    \approx
    2 \frac{\alpha}{v}
    \frac{\omega}{\sigma_\bot}
    \sqrt
    {
        \frac
        {
            r
        }
        {
            1- r
        }
    }
    \ .
\end{equation}
The second expression for $I_{\rm th}$ in (\ref{eq:Ith}) was obtained by replacing $\theta_{\rm th} \approx \theta_{\rm CS}$, which is valid at sufficiently low frequencies.
Note, that the threshold $I_{\rm th}$ vanish in the limit $\omega \rightarrow 0$.

To analyze the stability of the magnetization precession in the OOP-regime around the CS of magnetization we consider small deviations $\delta\theta$, $\delta\psi$ of angles $\theta$, $\psi$ from their stationary values $\theta_{s\,}$ and $\psi_{s\,}$, respectively. Using the standard technique of the stability analysis for linearized equations with $\delta\theta$, $\delta\psi$, we found the following two conditions of stability:
(i) $d \theta / d I_{\rm RF} > 0$, and
(ii) $6 \sin^{2}\theta > 4 + r - \sqrt{ 4 (2 - r)^2 - 3 r^2 }$.
The condition (i) determines the stable branch of solutions for $\theta$, i.e., the branch for which the angle $\theta$ increases with current magnitude $I_{\rm RF\,}$.
The condition (ii) is satisfied for $\theta = \theta_{\rm th}$ for any $0<r<1$ and, thus, is always satisfied on the increasing branch $d\theta/dI_{\rm RF}>0$.

The output DC voltage generated by an SD in the regime of stationary OOP magnetization precession can be evaluated as:
$U_{\rm DC} = \< I(t) R(t) \> = I_{\rm RF} R_\bot \< \cos(\omega t) / [1 + \tau \cos\beta(t)] \>$,
where
$R(t) = R_\bot/\left[1 + \tau \cos\beta(t)\right]$ is the MTJ resistance \cite{Miwa2014NatMater,Fang2019PRAppl},
$R_\bot$ is the junction resistance for $\beta = \pi/2$,
$\tau = {\rm TMR}/(2 + {\rm TMR})$,
${\rm TMR}$ is the tunneling magnetoresistance ratio of the MTJ,
$\cos\beta(t) = {\bm m}(t) \cdot {\bm p} = \sin\theta_s \cos(\omega t + \psi_s)$.
Calculating analytically $\< \cos(\omega t) / [1 + \tau \sin\theta_s \cos(\omega t + \psi_s)] \>$, and using previously given expression for $\cos\psi_s$ and assumption $\omega \approx \omega_p(\theta_s)$, the output DC voltage can be written in the following form:
\begin{equation}
\label{eq:Udc}
    U_{\rm DC}
    \approx
    2 \alpha
    \frac{w}{v}
    R_\bot
    \frac{\omega}{\sigma_\bot}
    \tan\theta_s
    \approx
    w I_{\rm th} R_\bot
    \, ,
\end{equation}
where 
$w \equiv w(a_\tau) = (1 - q_\tau)/a_\tau q_\tau$ is the dimensionless function of parameter $a_\tau = \tau \sin\theta_s$,
$q_\tau = \sqrt{1-a^2_\tau}$.

As one can see, close to the threshold ($I_{\rm RF} \gtrsim I_{\rm th\,}(\omega)$) the output DC voltage $U_{\rm DC}$ of a SD virtually does not depend on the input RF current magnitude $I_{\rm RF}$, and linearly increases with the frequency $\omega$.

For an SD with an average resistance $R_0$ the energy harvesting efficiency $\zeta$ can be defined as a ratio between the detector's output DC power $P_{\rm DC} = U^2_{\rm DC}/R_0$, and the power of input RF signal $P_{\rm RF} = I^2_{\rm RF} R_0/2$:
\begin{equation}
\label{eq:zeta}
    \zeta
    =
    \frac{
        P_{\rm DC}
    }
    {
        P_{\rm RF}
    }
    \approx
    2
    \left(\frac{1 - q_\tau}{a_\tau}\right)^2
    \left(\frac{I_{\rm th}}{I_{\rm RF}}\right)^2
    \, .
\end{equation}
For $I_{\rm RF} \ge I_{\rm th}$ the maximum value of $\zeta$ (achieved at the threshold) depends on the ${\rm TMR}$ ratio, $\zeta_{\rm max} \approx 2 (1 - q_\tau)^2/a^2_\tau$, and can reach $\zeta_{\rm max} \approx 40$\% for the ${\rm TMR}$ of 600\% experimentally achieved in \cite{Ikeda2008APL}. However, with a decrease of ${\rm TMR}$ the maximum value of $\zeta_{\rm max}$ is substantially reduced, and, for instance, for ${\rm TMR} = 1$ one can obtain only $\zeta_{\rm max} \approx 6$\%. Note, also, that in real experiments the measured values of $\zeta$ may be substantially smaller than $\zeta_{\rm max}$ value due to the impedance mismatch \cite{Pozar2012Book} between the  input transmission line with the impedance $Z_{\rm TL}$ and the SD with the impedance $Z_{\rm SD}$ connected to that line.
To account for this effect, one should use in (\ref{eq:zeta}) the effective input power delivered into the SD, $P_{\rm eff} = P_{\rm RF}(1 - |\Gamma|^2)$, instead of the incident power $P_{\rm RF}$, where $\Gamma = (Z_{\rm SD}-Z_{\rm TL})/(Z_{\rm SD}+Z_{\rm TL})$ is the complex reflection coefficient \cite{Pozar2012Book}.


To compare the results of the developed analytical theory to the experimental results,  and the results of numerical simulations, we consider the case of a SD based on a MTJ with the following parameters \cite{Fang2019PRAppl}: 
the FL of the thickness $l = 1.65$~nm has an elliptical cross section of $a \times b = 150 \times 50$~nm${}^2$;  normalized saturation magnetization of the FL $\mu_0 M_s = 1194$~mT, first order PMA field $B_1 = 1172$~mT,
second order PMA field $B_2 = 64$~mT, Gilbert damping constant $\alpha = 0.02$, the spin-polarization efficiency of the current is $\eta = 0.6$.
For simplicity, we assume $N_z = 1$, thus $N_x = N_y = 0$.
Then, the CS ratio is $r = (\mu_0 M_s - B_1)/B_2 = (\omega_M - \omega_1)/\omega_2 = 0.344$, and the equilibrium CS polar angle $\theta_{\rm CS} = \arccos \sqrt{1-r} \approx 36^\circ$.
Also, using the experimentally found values of the MTJ resistance in parallel ($R_P = 640 \ \Omega$) and antiparallel ($R_{AP} = 1236 \ \Omega$) states and tunneling magnetoresistance ratio ${\rm TMR} = (R_{AP}-R_P)/R_P = 0.93$, one can calculate the resistance of the SD in the perpendicular magnetic state ($\beta = \pi/2$) $R_\bot = 2 R_{AP} R_P / (R_{AP} + R_P) = 843 \ \Omega$.


\begin{figure}
\includegraphics{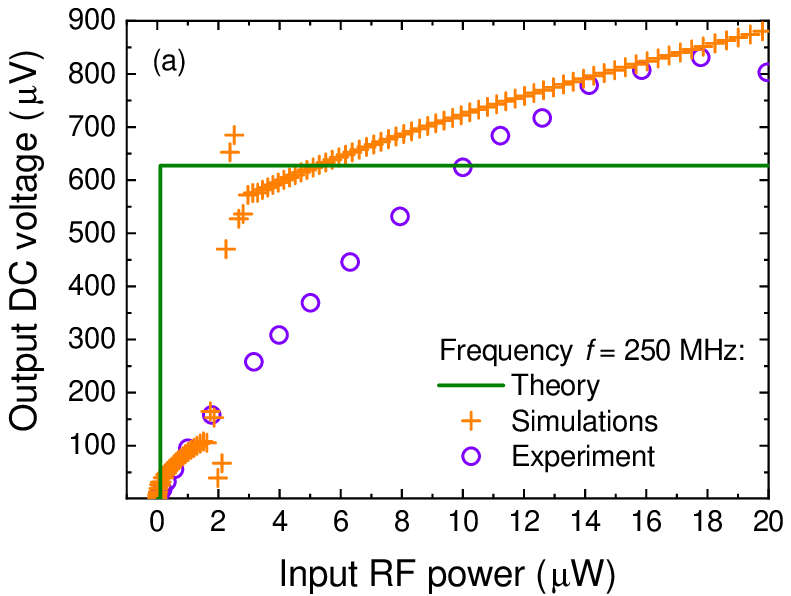}
\includegraphics{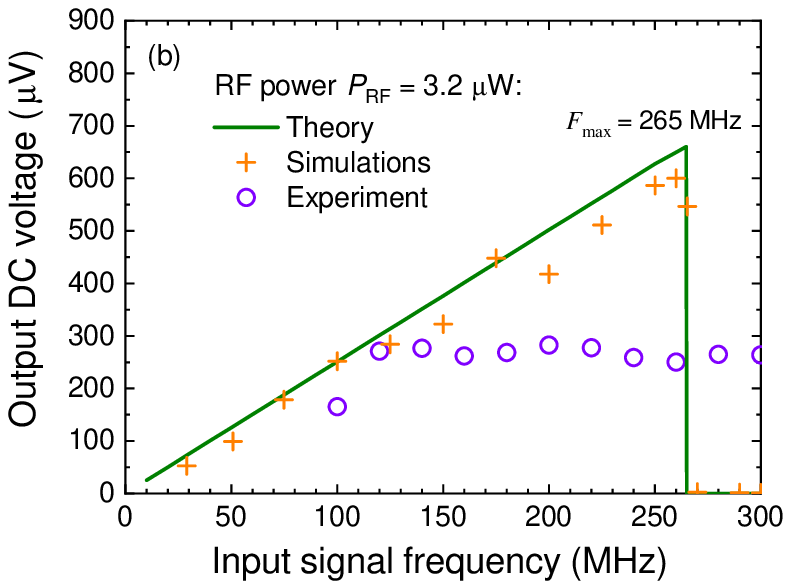}
\caption
{
(Color online)
The dependence of the output DC voltage $U_{\rm DC}$ of a SD with chosen typical parameters on:
(a) input signal power $P_{\rm RF} = I^2_{\rm RF} R_0/2$ for the signal frequency of $f = 250$~MHz;
(b) input signal frequency $f$ for the RF power $P_{\rm RF} = 3.2 \ \mu$W.
Green solid lines are the analytical dependences given by Eq.~(\ref{eq:Udc}).
Crosses show the results of numerical macrospin simulations.
Hollow violet circles correspond to the experimental data from \cite{Fang2019PRAppl}.
All the other parameters are presented in the text.
}
\label{f:Results}
\end{figure}

To check the validity of the developed analytical theory we performed macrospin simulations \cite{Macrospin} based on the numerical solution of (\ref{eq:LLGS}), and, then, numerically calculated the output DC voltage using the general expression $U_{\rm DC} = \< I(t) R(t) \>$. The results obtained from the developed analytical theory (green solid lines), from our simulations (crosses), and the experimental results from \cite{Fang2019PRAppl} (circles) are presented in Fig.~\ref{f:Results}. As one can see from the analytical expressions, the response of the SD to an input RF power $P_{\rm RF} = I^2_{\rm RF} R_0/2$ is non-zero and relatively weakly changes with $P_{\rm RF}$ for $P_{\rm RF}$ exceeding the frequency-dependent power threshold $P_{\rm th}(\omega) = I^2_{\rm th}(\omega) R_0/2$. Indeed, the analytical and numerical RF-power dependence of the output DC voltage $U_{\rm DC}$ has a step-like shape (see Fig.~\ref{f:Results}(a)): $U_{\rm DC} \approx 0$ for $P_{\rm RF} < P_{\rm th}$ and $U_{\rm DC} \approx {\rm const}$ for $P_{\rm RF} \ge P_{\rm th\,}$.
The non-resonant response of the considered SD to the variation of the  RF signal frequency $f$  can be clearly seen in Fig.~\ref{f:Results}(b). The output DC voltage of the SD obtained in both analytical theory and numerical simulations increases linearly with $f$  for $f < f_{\rm th}$, and  vanishes ($U_{\rm DC} \approx 0$) when the RF signal frequency exceeds a certain threshold  $f \ge f_{\rm th}$.

The existence of the threshold frequency $f_{\rm th}$ follows from Eq.~(\ref{eq:Ith}): with an increase of the signal frequency $f$, the threshold power $P_{\rm th}$ required for the  proper SD operation also increases, and at the point where  this threshold $P_{\rm th}$ exceeds the input power $P_{\rm RF}$, the magnetization oscillations in the device FL vanish. Thus, as expected, in the OOP-regime the SD works as a broadband non-resonant threshold RF detector.

It should be noted, that, while the analytically and numerically calculated dependences of the output DC voltage $U_{\rm DC}$ on the input signal power $P_{\rm RF}$ and frequency $f$, presented in Fig.~\ref{f:Results}, are in reasonable agreement, these dependencies demonstrate only qualitative resemblance with the experimental results from \cite{Fang2019PRAppl}.
We believe, that this discrepancy between the experiment \cite{Fang2019PRAppl} and theory could be explained by an influence of the in-plane anisotropy in the FL of the SD  used in the experiment \cite{Fang2019PRAppl}, which was not taken into account in our theoretical model, and, also, by the possible excitation in the experiment of some transitional regimes of the magnetization precession at rather large values of $P_{\rm RF}$ and $f$. Both these effects require an additional theoretical and experimental study. At the same time, we note that the presented theory, nonetheless, allowed us to approximately evaluate the experimentally obtained $U_{\rm DC}$ in the interval of variation of both $P_{\rm RF})$ and the signal frequency $f$.
Also, it is important to note, that the efficiency $\zeta$ of the RF/DC energy conversion (or RF energy harvesting) for a SD with chosen typical parameters used in our numerical simulations reaches $5.7$\% at the threshold, which well agrees with the analytical estimation from (\ref{eq:zeta}). This number is rather encouraging for the possible use of SD having perpendicular anisotropy of the FL in practical RF energy harvesting.  If we take into account the above discussed effect of impedance mismatch, which reduces the efficiency $\zeta$ to $1.5$\%, even this last lower number for the energy harvesting efficiency could be sufficient for many practical applications.

\begin{figure}
\includegraphics{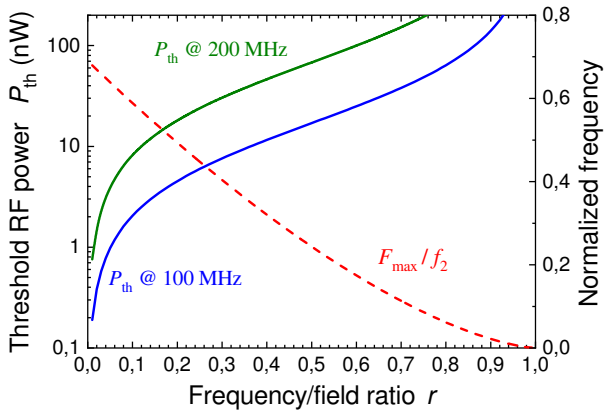}
\caption
{
(Color online)
Numerically calculated threshold RF power $P_{\rm th}$ (left axis; calculated from (\ref{eq:Ith}) at frequencies 100~MHz and 200~MHz) and normalized frequency ${\rm max}\{ \omega_p\} / \omega_2$ (right axis) as a function of the CS ratio $r$ for a SD with typical parameters.
}
\label{f:r}
\end{figure}

It  should be noted, that the analytical theory developed in this work could be very important for the optimization of working characteristics of SD-based RF energy harvesters. In particular,  it follows from the Eqs.~(\ref{eq:Ith}) and (\ref{eq:Udc}) that such important SD performance parameters as RF power threshold $P_{\rm th}$ and the maximum operational frequency $f_{\rm th}$ strongly depend on the CS ratio $r = (\mu_0 M_s - B_1) / B_2$ (written here for the case $N_z = 1$). In experiments the effective PMA field $B_2$ can be controlled by the variation of the FL thickness \cite{Ikeda2008APL,Amiri2011APL}, which allows one to vary the ratio $r$ in a rather wide range.
As one can see from Fig.~\ref{f:r} and (\ref{eq:Ith}), with the increase of the ratio $r$ the power threshold $P_{\rm th}$ of energy harvesting strongly increases, while the maximum operational frequency  $f_{\rm th}$ linearly decreases. Thus, it is preferable to work at low values of the ratio $r$, which corresponds to very  small thicknesses of the SD FL that are difficult to achieve experimentally \cite{Ikeda2008APL,Amiri2011APL}.
Thus, a compromise is necessary, and Eqs.~(\ref{eq:Ith}) and (\ref{eq:Udc}) allow one to find that at the experimentally reachable values of $r \approx 0.2 - 0.4$ it is possible to achieve reasonably small values of the threshold power $P_{\rm th} \lesssim 20$~nW, while keeping the maximum operational frequency around 200- 250~MHz (see Fig.~\ref{f:r}).


In conclusion, we have shown theoretically that a spintronic diode (SD) having a first and second order perpendicular magnetic anisotropy of the free layer can be used as an efficient RF signal detector and energy harvester operating in the absence of a bias magnetic field. The device generates a finite output DC voltage, when its input RF power $P_{\rm RF}$ exceeds a certain frequency-dependent threshold $P_{\rm th}$, while at $P_{\rm RF} > P_{\rm th}$ the voltage weakly depends on power. Such a regime of diode operation is possible at sufficiently low frequencies below the threshold frequency dependent on the RF signal power and CS ratio $r$, which has an optimal range of values $r \approx 0.2-0.4$.
Finally, it was demonstrated that the energy harvesting efficiency for the harvester could exceed 5\% ($1.5$\% with an account of the impedance mismatch effect) that is sufficient for some RF energy harvesting applications.

Acknowledgements

This work was supported in part by the U.S. National Science Foundation (Grant No. EFMA-1641989), by the U.S. Air Force Office of Scientific Research under MURI Grant No. FA9550-19-1-0307, and by the Oakland University Foundation.
This work was also supported in part by the grant No.~19BF052--01 from the Ministry of Education and Science of Ukraine and by the NATO SPS.MYP grant No.~G5792.



\begin{thebibliography}{99}

\bibitem{Harb2011RenEn}
A.~Harb,
{Renew. Energy} \textbf{36}, 2641 (2011).

\bibitem{Duroc2014Book}
Y.~Duroc, G.~Andia Vera,
Towards Autonomous Wireless Sensors: RFID and Energy Harvesting Solutions,
in: S.C.~Mukhopadhyay (Ed.),
{\emph{Internet of Things: Challenges and Opportunities}
(Smart Sensors, Measurement and Instrumentation, Vol.~9)}
(Springer, Berlin, 2014).

\bibitem{Valenta2014MMag}
C.R.~Valenta and G.D.~Durgin,
{IEEE Microw. Mag.} \textbf{15}, 108 (2014).

\bibitem{Divakaran2019MCAE}
S.K.~Divakaran, D.D.~Krishna, Nasimuddin,
{Int. J. RF Microw. Comput. Aided Eng.} \textbf{29}, e21633 (2019).

\bibitem{Pozo2019Electronics}
B.~Pozo, J.I.~Garate, J.\'{A}.~Araujo and S.~Ferreiro,
{Electronics} \textbf{8}, 486 (2019).

\bibitem{Tulapurkar2005Nature}
A.A.~Tulapurkar, Y.~Suzuki, A.~Fukushima \emph{et al.},
{Nature} \textbf{438}, 339 (2005).

\bibitem{Prokopenko2013Book}
O.V.~Prokopenko, I.N.~Krivorotov, T.J.~Meitzler \emph{et al.},
Spin-Torque Microwave Detectors,
in: S.O.~Demokritov and A.N. Slavin (Eds.),
{\emph{Magnonics: From Fundamentals to Applications}
(Topics in Applied Physics, Vol.~125)}
(Springer, Berlin, 2013).

\bibitem{Prokopenko2015LTP}
O.V.~Prokopenko and A.N.~Slavin,
{Low Temp. Phys.} \textbf{41}, 353 (2015).

\bibitem{Ishibashi2010APEx}
S.~Ishibashi, T.~Seki, T.~Nozaki \emph{et al.},
{Appl. Phys. Express} \textbf{3}, 073001 (2010).

\bibitem{Prokopenko2011APL}
O.~Prokopenko, G.~Melkov, E.~Bankowski \emph{et al.},
{Appl. Phys. Lett.} \textbf{99}, 032507 (2011).

\bibitem{Miwa2014NatMater}
S.~Miwa, S.~Ishibashi, H.~Tomita \emph{et al.},
{Nature Mater.} \textbf{13}, 50 (2014).

\bibitem{Fang2016NatCommun}
B.~Fang, M.~Carpentieri, X.~Hao \emph{et al.},
{Nature Commun.} \textbf{7}, 11259 (2016).

\bibitem{Cheng2010PRL}
X.~Cheng, C.T.~Boone, J.~Zhu, and I.N.~Krivorotov,
{Phys. Rev. Lett.} \textbf{105}, 047202 (2010).

\bibitem{Prokopenko2012JAP}
O.V.~Prokopenko, I.N.~Krivorotov, E.~Bankowski \emph{et al.},
{J. Appl. Phys.} \textbf{111}, 123904 (2012).

\bibitem{Fang2019PRAppl}
B.~Fang, M.~Carpentieri, S.~Louis \emph{et al.},
{Phys. Rev. Appl.} \textbf{11}, 014022 (2019).

\bibitem{Tomasello2020PRAppl}
R.~Tomasello, B.~Fang, P.~Artemchuk \emph{et al.},
{Phys. Rev. Appl.} \textbf{14}, 024043 (2020).

\bibitem{Artemchuk2020Book}
P.Yu.~Artemchuk and O.V. Prokopenko,
Detection of Microwave and Terahertz-Frequency Signals in Spintronic Nanostructures,
in: A.~Kaidatzis, S. Sidorenko, I. Vladymyrskyi and D. Niarchos (Eds.),
{\emph{Modern Magnetic and Spintronic Materials: Properties and Applications}
(NATO Science for Peace and Security Series B: Physics and Biophysics)}
(Springer, Dodrecht, 2020).

\bibitem{Macrospin}
In \cite{Fang2019PRAppl} it was shown that the SD's FL has a single domain magnetic state, so the use of macrospin approximation is acceptable.

\bibitem{Melkov1996Book}
A.G.~Gurevich and G.A.~Melkov,
\emph{Magnetization Oscillations and Waves}
(CRC Press, New York, 1996).

\bibitem{Slavin2009IEEETMagn}
A.~Slavin and V.~Tiberkevich,
{IEEE Trans. Magn.} \textbf{45}, 1875 (2009).

\bibitem{Ikeda2008APL}
S.~Ikeda, J.~Hayakawa, Y.~Ashizawa \emph{et al.},
{Appl. Phys. Lett.} \textbf{93}, 082508 (2008).

\bibitem{Pozar2012Book}
D.M.~Pozar,
{\emph{Microwave Engineering}, fourth ed.}
(John Wiley \& Sons, New York, 2012).


\bibitem{Amiri2011APL}
P.~Khalili Amiri, Z.M.~Zeng, J.~Langer \emph{et al.},
{Appl. Phys. Lett.} \textbf{98}, 112507 (2011).


\end{thebibliography}
\end{document}